\documentclass{article}
\usepackage[T1]{fontenc} %
\usepackage[utf8x]{inputenc} %
\usepackage{ismir,amsmath,cite,url}
\usepackage{amssymb}
\usepackage{graphicx}
\usepackage{xcolor}
\usepackage{enumitem}

\usepackage{epigraph}
\usepackage{lineno}
\usepackage{lipsum}
\usepackage{csquotes}
\usepackage{xspace}
\usepackage{pgfplots}
\usepackage{balance}
\usepackage{lastpage}

\newcommand{\jaidev}[1]{{\color{red}{\textbf{Jaidev}: #1}}}

\newcommand{\etal}{\emph{et al.}}
\newcommand{\cf}{\emph{cf.}}

\newcommand{\newpara}[1]{\vspace{1mm}\noindent\textbf{#1}}

\newcommand{\mcS}{\mathcal{S}}
\newcommand{\mcB}{\mathcal{B}}
\newcommand{\mcV}{\mathcal{V}}
\newcommand{\mcM}{\mathcal{M}}

\definecolor{green}{RGB}{62, 250, 63}

\newcommand\harrypotterfont[1]{{\usefont{T1}{harrypotter}{m}{n} #1 }}

\nolinenumbers

\makeatletter
\DeclareRobustCommand\onedot{\futurelet\@let@token\@onedot}
\def\@onedot{\ifx\@let@token.\else.\null\fi\xspace}

\def\eg{\emph{e.g}\onedot} 
\def\ie{\emph{i.e}\onedot} 
\def\cf{\emph{cf}\onedot}

\def\etal{\emph{et al}\onedot}
\makeatother

\title{{ {\LARGE \harrypotterfont{Sonus~Texere!}}} Automated Dense Soundtrack Construction for Books using Movie Adaptations}

\newcommand{\copyrightpapername}{Sonus~Texere! Automated Dense Soundtrack Construction for Books using Movie Adaptations}

\multauthor
  {Jaidev Shriram \hspace{1cm} Makarand Tapaswi \hspace{1cm} Vinoo Alluri} 
  {International Institute of Information Technology, Hyderabad \\
  \small{\url{https://auto-book-soundtrack.github.io/}} \\
  \footnotesize{\tt jaidev.shriram@students.iiit.ac.in, makarand.tapaswi@iiit.ac.in, vinoo.alluri@iiit.ac.in}
\vspace{-4mm}
}

\def\authorname{J. Shriram, M. Tapaswi, and V. Alluri}

\usepackage[colorlinks=true,pagebackref=true,bookmarks=false,pdfauthor={\authorname},pdfsubject={\papersubject}]{hyperref}

\begin{document}

\maketitle

\begin{abstract}

Reading, much like music listening, is an immersive experience that transports readers while taking them on an emotional journey.
Listening to complementary music has the potential to amplify the reading experience, especially when the music is stylistically cohesive and emotionally relevant.
In this paper, we propose the first fully automatic method to build a dense soundtrack for books, which can play high-quality instrumental music for the entirety of the reading duration.
Our work employs a unique text processing and music weaving pipeline that determines the context and emotional composition of scenes in a chapter.
This allows our method to identify and play relevant excerpts from the soundtrack of the book's movie adaptation.
By relying on the movie composer's craftsmanship, our book soundtracks include expert-made motifs and other scene-specific musical characteristics.
We validate the design decisions of our approach through a perceptual study.
Our readers note that the book soundtrack greatly enhanced their reading experience, due to high immersiveness granted via uninterrupted and style-consistent music, and a heightened emotional state attained via high precision emotion and scene context recognition.

\end{abstract}

\section{Introduction}\label{sec:introduction}

\begin{figure}[t]
\centering
\includegraphics[width=0.7\linewidth]{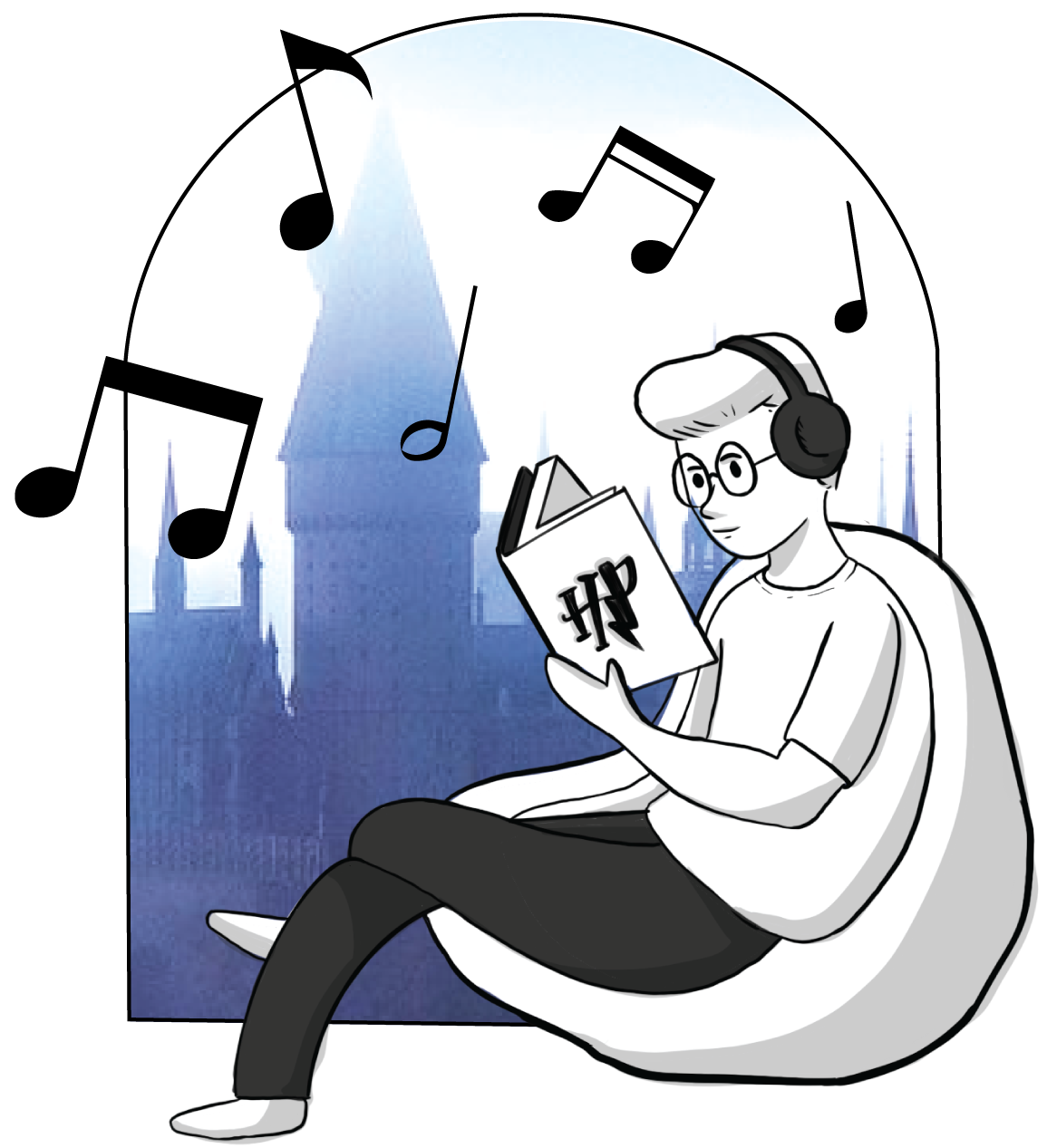}
\vspace{-2mm}
\caption{We aim to transport readers to a musical universe by building a dense and coherent book soundtrack that is borrowed from movie adaptations.} 
\label{fig:teaser}
\vspace{-4mm}
\end{figure}

In 1975, a short repetitive piece of orchestral music gained notoriety for inducing fear in its listeners, putting them in a suspenseful state of impending doom.
Today, it still induces the same reaction, right when the iconic shark of the movie \textit{Jaws} appears on screen.
Movie composers, such as John Williams who created the \textit{Jaws} theme, have long been aware of the impact music has on its listeners.
Well-placed music can accentuate scenes to raise the emotional stakes or foreshadow upcoming events; ill-suited music can even suggest that something evil is afoot~\cite{Boltz2001MusicalSA}.
When taken as a whole, the musical imagery afforded by such soundtracks greatly complements the movie watching experience.
In this paper, we attempt to answer whether we can create a similar experience for books (\cf~Fig.~\ref{fig:teaser}).

Not unlike movies, reading literature can be an incredibly transportive process that puts one in a meditative state, while actively engaging their imagination and mind~\cite{pennington2017reading}.
Apart from the lack of visuals, books share many similarities from a music composer's perspective: they have recurring themes, characters (allowing for unique \emph{leitmotifs}~\cite{ChekowskaZacharewicz2020MusicEA}), and even long-drawn emotional and narrative arcs that can be teased and reinforced musically.
Still, it is uncommon to see soundtracks that are tailored for books; it is up to the reader to curate their own playlist and craft their experience.
Inconvenience aside, this requires the reader to preempt the type of music expected for a book’s chapter and switch songs at relevant plot points. This is simply infeasible.
In this work, we resolve these issues by proposing the first automatic system to build a dense soundtrack that plays throughout the entire reading duration of a book.

Specifically, we focus on building a soundtrack for books with movie adaptations.
This allows us to draw on the corresponding movie soundtrack and take advantage of the composer/director's musical intents and instincts for the \emph{same story}.
However, adapting the soundtrack is far from trivial due to missing alignment between book parts and music segments.
Our approach resolves this by finding scenes in the movie adaptation that match parts of the book and searching for music that plays in that movie scene.
This results in high quality, narrative specific matches that amplify the reading experience. 
We conduct our experiments on
\textit{Harry Potter and the Philosopher's Stone}, chosen due to its popularity and Academy Award nominated soundtrack.
We evaluate the quality of our soundtrack by asking participants to read two music-accompanied chapters of the book followed by a semi-structured interview.

Our contributions can be summarised as follows:
(i) We propose the first fully automatic approach for constructing book-long soundtracks.
(ii) Our pipeline matches the story in the book and movie to lift musical cues from the movie adaptation. Gaps in this alignment are filled through an emotion-driven music retrieval system.
(iii) A perceptual study of the soundtrack validates our proposed approach and provides further insights into soundtracking for books.

\section{Related Work}

\newpara{Soundtracking for books.}
Methods for constructing book (or story) soundtracks can be divided into two: generative or retrieval based.
Generative approaches typically parse the text for concepts and provide that as input to a music generation pipeline.
Topic extraction followed by sentiment analysis~\cite{Salas2018GeneratingMF} or density estimation of emotion-related words~\cite{Davis2014GeneratingMF} are used to generate melodies.
On the other hand, retrieval based approaches, mine text for semantic concepts that are used to retrieve ambient music (sometimes with pitch correction)~\cite{Harmon2017NarrativeinspiredGO} or apply similar ideas to Twitter texts~\cite{Thorogood2013ComputationallyCS}.
The general idea of cross-modal text-to-audio retrieval~\cite{Oncescu2021AudioRW} is adapted for tag-based music~\cite{Won2021MultimodalML} or characterises documents and music as a distribution over emotions~\cite{Cai2007MusicSenseCM}.
Recently, Won~\etal~\cite{won2021emotion} propose a joint emotion-driven embedding space for story sentences and music enabling cross-modal retrieval.
However, understanding emotions conveyed in a book depends on the larger narrative requiring a context of more than a few sentences.
Thus, while \cite{won2021emotion}~could be used to construct dense book soundtracks in theory, the potential switching of music and resulting change in emotion every few sentences may not result in a good user experience.

Unlike above approaches, we focus on retrieving music from the soundtrack of a book's movie adaptation.
Our approach assembles coherent music for books by considering large chunks of the book chapter, aligning them with movie scenes, and finding matched music pairs.
We fill in the gaps (only unmatched chapter segments) by relying on emotion-based matching.
To the best of our knowledge, we are the first to perform narrative-specific music retrieval and \emph{weave} a soundtrack for the entire book.

\newpara{Multi-modal music recommendation}
systems use cues such as user location, time, and environmental information to play music in day-to-day life~\cite{Reddy2006LifetrakMI} or even traffic conditions for music in cars~\cite{Baltrunas2011InCarMusicCM}.
Perhaps closest to our work, PICASSO is a fine example of a ranking model trained on pairs of matching music and movie-clips (images, subtitles) that is used to provide music recommendations for image slideshows or audio books~\cite{Stupar2011PICASSO}.
Instead, our work focuses on producing a \emph{dense} book soundtrack by aligning narrative components of the book with the movie adaptation.
This results in high precision narrative matches and ensures a coherent soundtrack adapted from the same story.

\begin{figure*}[t]
\centering
\includegraphics[width=0.9\linewidth]{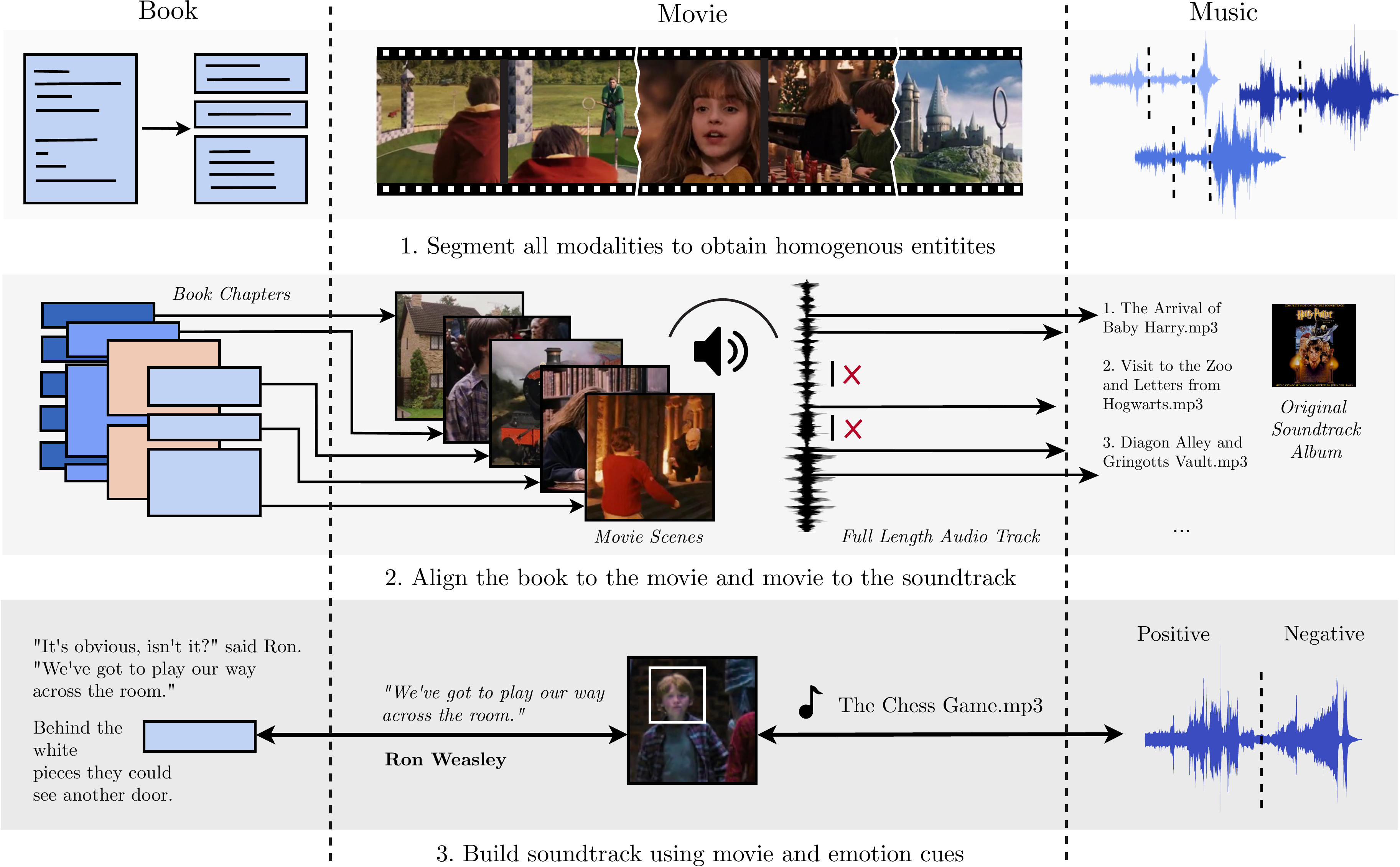}
\caption{\textbf{Overview of approach.} We present a novel technique to build a musically rich book-length soundtrack. We accomplish this by first segmenting the book, its adaptation, and music into smaller homogenous chunks. These segments are then matched with the movie acting as an intermediary for text and music, thereby producing the final soundtrack.}
\label{fig:approach_overview}
\end{figure*}

\section{Assembling Soundtracks for Books}

A large book may take several (variable) hours for a reader to complete.
Within the book, there are multiple chapters, each having sections indicated by the theme, emotion, or location.
An ideal soundtrack should respect these scene boundaries and change accordingly.
On the music front, while different tracks from the soundtrack are typically homogeneous, they are rarely played in their entirety in the movie at a single stretch. A track may include the score for the setup, conflict, climactic moment, and resolution for a certain storyline, which composers will selectively play at opportune moments to maximise emotional payoffs.

We use the movie adaptation as an intermediary between the two that links plot points from the book with snippets of music from the soundtrack album.
We first divide each modality - the book, music soundtrack, and the movie into smaller segments (Sec.~\ref{subsec:segmentation}).
Then, we obtain clues about the soundtrack associated with each plot point in the book by aligning both the book and soundtrack to the movie (Sec.~\ref{subsec:book_movie_music}).
Finally, we weave together the music for the book by combining movie-based and emotion-based matches (Sec.~\ref{subsec:book_music}).
Fig.~\ref{fig:approach_overview} illustrates this overall flow.

\subsection{Segmenting the Book, Music, and Movie}
\label{subsec:segmentation}

We discuss methods for segmenting the book chapters $\mcB_i$ in a book $\mcB = [\mcB_1, \ldots, \mcB_L]$;
identifying cohesive musical segments within each track $M_j$ of the album $\mcM = [M_1, \ldots, M_P]$;
and identifying scene boundaries for a movie $\mcV = [V_1, \ldots, V_Q]$ that facilitates the alignment.

\newpara{Book narrative segmentation.}
We divide the text in a book chapter based on narrative-relevant factors such as theme, location, activity composition, character constellation, or even time~\cite{Zehe2021DetectingSI}.
Due to the absence of large datasets for this task, we adopt an unsupervised approach, temporally weighted hierarchical clustering (TW-FINCH)~\cite{sarfraz2021temporally}, recently shown to be successful on video activity segmentation.
For a chapter $\mcB_i$, we encode each paragraph using a pretrained language model $\phi_{LM}(\cdot)$ (specifically MPNet~\cite{song2020mpnet} that performs well for sentence embeddings and semantic search~\cite{reimers-2019-sentence-bert}), and cluster semantically similar and spatially close paragraphs to produce disjoint partitions,
\begin{equation}
\{\mcB_i^p\}_{p=1}^{K_i} = \text{TW-FINCH}\left( \phi_{LM}(\mcB_i) ) \right) \, .
\end{equation}
We align individual segments $\mcB_i^p$ with music segments.

We also considered a few baselines but ignored them due to inferior results.
TextTiling~\cite{Hearst1997TextTS} tended to uniformly partition chapters while TopicTiling~\cite{Riedl_topictil} often resulted in over-segmentation.
In contrast, our approach yielded segments that mostly respected narrative shifts.

\newpara{Keystrength based music segmentation.}
As mentioned earlier, the entire track $M_j$ from an album is (almost) never played in a portion of the movie.
Hence, similar to book chapters, we focus on creating homogeneous emotionally-cohesive track segments that can be played continuously.

\begin{figure*}[t]
\centering
\includegraphics[width=0.9\linewidth]{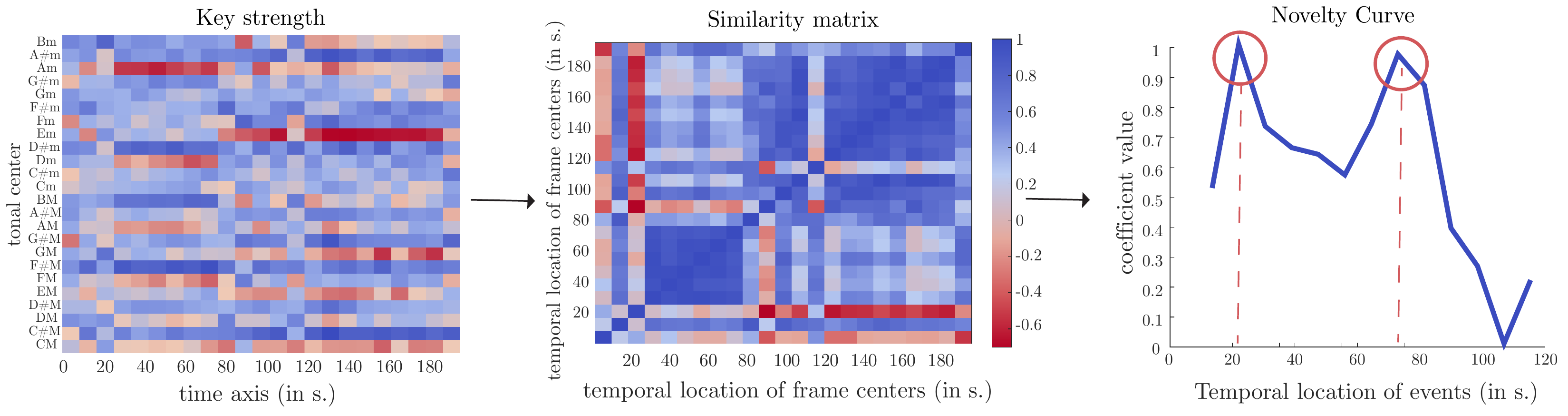}
\caption{\textbf{Music segmentation pipeline.} We segment all tracks from the soundtrack to ensure a cohesive listening experience.
We extract keystrength~\cite{Gmez2006TonalDO} that captures tonal properties of a soundtrack (left), compute the self-similarity matrix (center), and use that to calculate the novelty curve \cite{Foote2003MediaSU}.
The peaks of this curve are used to segment the track.}
\label{fig:music_segment}
\end{figure*}

A key feature that music segments are required to reflect are emotions that the director/composer intends to convey.
Since we expect homogeneity in music segment emotions, \emph{keystrength}~\cite{Gmez2006TonalDO} is a good attribute as it captures the tonal properties of music, and tonal changes likely suggest emotional shifts (\eg~major to minor mode suggesting a potential shift from positive to negative valence~\cite{Juslin2011HandbookOM}).
\textit{Keystrength} as extracted via the MIRToolbox~\cite{Lartillot2008} gives us a probability for each possible key resulting in a 24 dimensional vector (12 major and 12 minor keys).
We use a window size of 10 seconds with an 85\% overlap, that helps ignore minor local variations in the track while retaining larger shifts.
We then compute a self-similarity matrix~\cite{Foote2003MediaSU} of the time-varying keystrength vector which is then used to calculate the novelty curve, with a kernel size of 64~\cite{Foote2003MediaSU}.
The peaks in the novelty curve determine points of change.
Finally, we use those peaks as our segment boundaries to produce $L_j$ music segments $\{ M_j^c \}_{c=1}^{L_j}$ (see Fig.~\ref{fig:music_segment}).

\newpara{Movie scene detection.}
As a third step, we segment the movie into narrative-coherent scenes $[V_1, \ldots, V_Q]$ using the approach described in~\cite{tapaswi_storygraph}.
First, the movie is divided into shots (consecutive video frames from the same camera).
Then, a dynamic programming algorithm is used to find scene boundaries (a scene consists of multiple shots) so as to maximise intra-scene similarity.

Summarising, we denote $\mcB_i^p$ as the segment of book chapter $\mcB_i$; $M_j^c$ as the music segment from track $M_j$; and $V_q$ as the $q^\text{th}$ scene from the movie $\mcV$.

\subsection{Aligning the Book, Movie, and Music}
\label{subsec:book_movie_music}

We first align the book with the movie adaptation using a new two-stage coarse-to-fine alignment scheme.
Then, we also align the movie audio to the soundtrack album.

\newpara{Chapter-scene coarse alignment.}
We first assign a set of scenes from the movie to each book chapter.
We use an approach similar to~\cite{TapaswiB2M} where pairwise similarities are computed between a book chapter $\mcB_i$ and a video scene $V_q$ via character histograms and matched dialogues.
The chapter-scene relationship is then encoded as a graph (each node represents a chapter-scene pair), with edge weights representing similarity scores.
Calculating the shortest path over this graph provides the alignment between all chapters and scenes.
Additional details are provided in the Appendix.

\begin{figure}[b!]
\centering
\includegraphics[width=\linewidth]{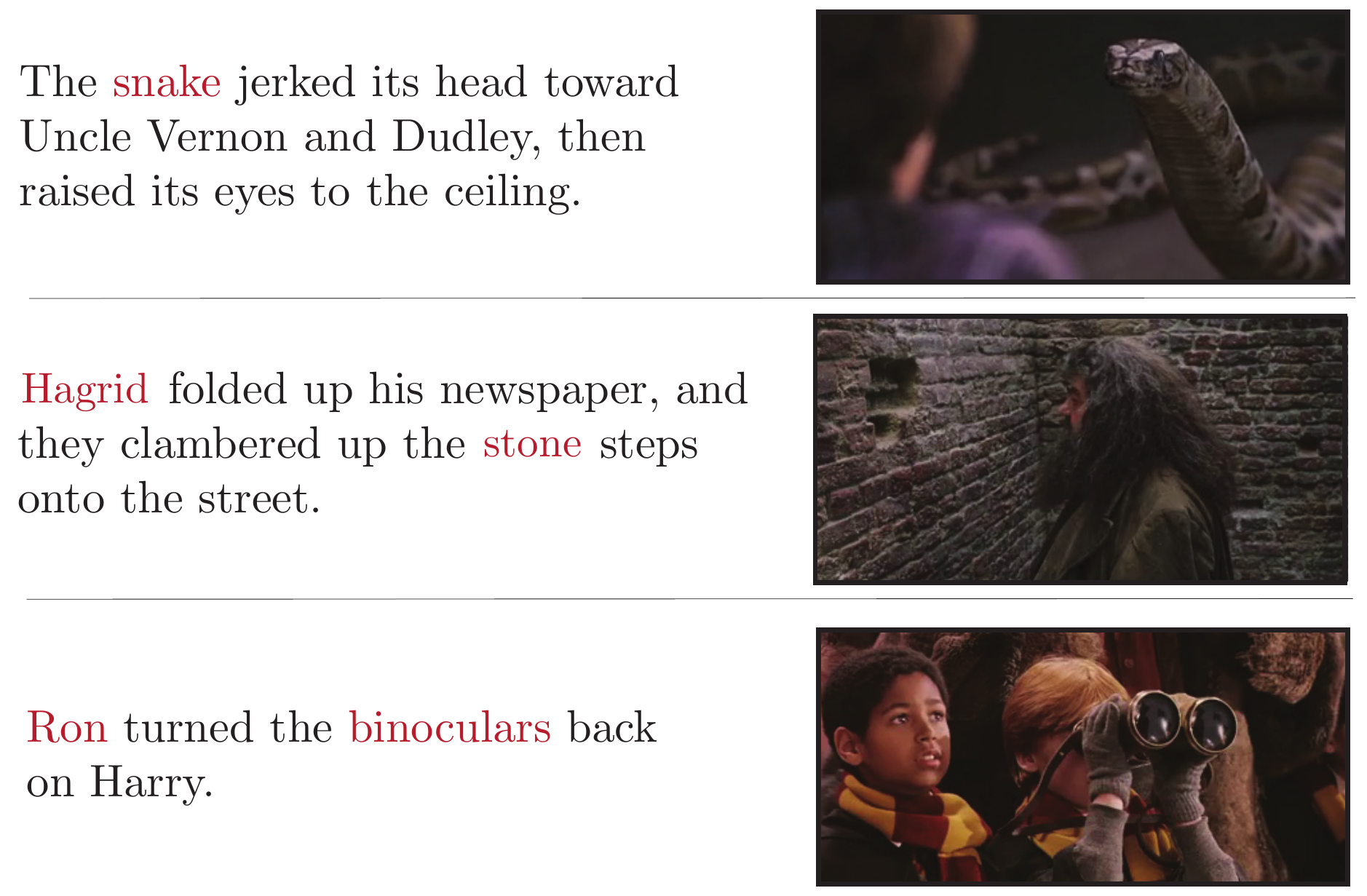}
\caption{\textbf{CLIP-based paragraph-scene alignment.}
Aligned examples of automatically selected visual sentences and their video frames in the scene using the vision-language model CLIP~\cite{radford2021learning} in a zero-shot setting.
Text highlights are words that we think that caused the match.}
\label{fig:clip}
\end{figure}

\newpara{Paragraph-scene refinement.}
The coarse alignment cannot be used directly for soundtracking as a chapter $\mcB_i$ contains distinct segments $\mcB_i^p$ that likely need different music.
Thus, in addition to sparse dialogue matches, we compute similarities between sentences in the chapter segment $\mcB_i^p$ and frames of the video scene $V_q$ using
a pretrained vision-language model (CLIP~\cite{radford2021learning}).
To improve the quality of CLIP matching scores, we prune dialog and mundane sentences from the chapter segment using a TF$\cdot$IDF based scoring system.
This emphasises relatively rare characters and objects that are likely to give stronger matches than commonly occurring ones.
We also retain sentences with a high \emph{concreteness} index~\cite{Brysbaert2014ConcretenessRF} that measures how likely a word can be seen or experienced (in contrast to \emph{abstract} words).
Finally, we take the top remaining sentences, encode them with CLIP, and calculate cosine similarity with all CLIP encoded video shot frames in the chapter's scenes (see Fig.~\ref{fig:clip}).
Scenes with a score higher than $\theta$ are assigned to the text segment using a mapping function,
\begin{equation}
\label{eq:map_book_movie}
A(\mcB_i^p) = \{ V_q : \text{CLIP}(\mcB_i^p, V_q) > \theta \} \, .
\end{equation}

\newpara{Aligning the movie audio track with the soundtrack album.}
We identify the music in the movie by passing its audio track through Shazam\footnote{\url{https://www.shazam.com/}}, a popular commercial audio-fingerprint based search application~\cite{Wang2003AnIS} using its free public API.
This results in high-precision estimates for the music played every few seconds.
While any audio fingerprinting approach would perhaps work, we found that Shazam was quite accurate in the presence of background noise and dialogue, which is pervasive in movies.

\subsection{Weaving the Book Soundtrack}
\label{subsec:book_music}

Post alignment with the movie, book chapter segments $\mcB_i^p$ belong to one of two sets:
(i) those associated with at least one movie scene $\mcS = \{ \mcB_i^p : |A(\mcB_i^p)| \geq 1 \}$; and
(ii) those without, $\bar{\mcS} = \{ \mcB_i^p : A(\mcB_i^p) = \varnothing \}$.
Recall, $A(\mcB_i^p)$ is the set of aligned video scenes to a book segment (\cf~Eq.~\ref{eq:map_book_movie}).

\newpara{Extracting emotion labels.}
For all text segments $\mcB_i^p$, we classify each paragraph into positive, neutral, or negative using a BERT-based emotion classifier $\phi_\text{BERT}$, trained on Reddit comments~\cite{demszky2020goemotions}.
A majority vote across paragraphs is used to assign the emotion label to the chapter segment $E_\text{book}(\mcB_i^p) = \text{mode}(\phi_\text{BERT}(\mcB_i^p))$.
For a music segment, similar to~\cite{Davis2014GeneratingMF}, we encode it's emotion as valence, $E_\text{music}(M_j^c)$, based on the mode of the song (major or minor).
This is based on literature that indicates that tracks in minor tend to be associated with negative emotions and tracks in major with positive emotions~\cite{Juslin2011HandbookOM}.
We also tried approaches that predict emotion from audio~\cite{won2021emotion,Eerola2009PredictionOM}, but they didn't work as well for our application.

\newpara{Importing music snippets from the video scene.}
For every text segment $\mcB_i^p \in \mcS$, we extract the movie timestamps corresponding to the matched dialogues or CLIP-based frame-sentence pairs.
We use the audio-search to identify the overall track $M_j$ being played at any of the above timestamps in the movie (in a small neighbourhood).
A specific music segment $M_j^c$ is chosen by matching emotion predictions \ie~$E_\text{book}(\mcB_i^p) = E_\text{music}(M_j^c)$ (we pick one randomly if there are several segments).
While we can pick emotion-matching music segments without the book-movie alignment,
it will likely result in spurious matches.

\newpara{Emotion-based retrieval.}
For the chapter segments that are not aligned with any video scene, $\bar{\mcS}$, and those in $\mcS$ that did not find an emotion compatible soundtrack, we assign a random music segment among the set of emotionally compatible compositions.
Note that while we can pick any music segment (even from different movies), we restrict to the soundtracks for this movie maintaining the composer's stylistic coherence.

\section{Results and Evaluation}

\begin{figure}[t]
\centering
\resizebox{\linewidth}{!}{\input{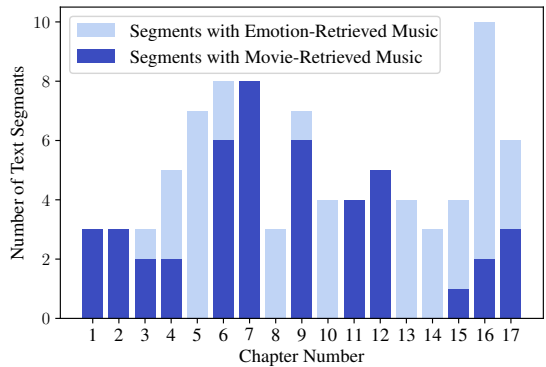}}
\caption{\textbf{Number of chapter segments with movie- vs. emotion-retrieved music.} Several chapters in the book are well represented in the movie and are predominantly soundtracked using movie cues. For others, we use an emotion-based retrieval method to build a soundtrack.}  
\label{fig:movie_music}
\end{figure}

Our approach is designed to be applicable to books that have mostly faithful movie adaptations in terms of few matching dialogues, a relatively similar plot, and at least some matching characters.
We evaluate it on the first book and movie pair in the \textit{Harry Potter} series, \textit{Harry Potter and the Philosopher’s Stone}.

\subsection{Harry Potter: A Case Study}
Our unsupervised text segmentation approach splits 17 chapters into 87 segments, with an average of 5.11 segments per chapter.
Assuming a reading speed of 250 words per minute, each segment requires a \textasciitilde4 minute track, for a total soundtrack duration of \textasciitilde6 hours.
Music segmentation of 19 tracks from the soundtrack album results in 47 audio chunks with an average duration of \textasciitilde52 seconds.
For the movie, we obtain 120 scenes using the scene detection algorithm.
Computing the book-movie alignment results in a strong chapter-scene alignment, with 82\% accuracy, calculated as the percentage of movie shots correctly assigned to a chapter~\cite{TapaswiB2M}.

For the fine-grained alignment, 34 chapter segments have dialogue matches, while the CLIP based refinement results in a 130\% improvement in the number of chapter segments associated with the movie; only 9 segments remain unmatched.
After performing the final book-music alignment, 45 chapter segments have music imported from a matching video scene through the book-movie alignment.
The remainder 42 segments are soundtracked using music segments fetched via emotion matching.
See Fig.~\ref{fig:movie_music} for a chapter-wise distribution.

\subsection{Experiment Design}

As the primary goal of our method is to improve the reading experience, we obtain user feedback by setting up semi-structured interviews with 10 individuals, who each read two chapters of the \textit{Harry Potter} book.
We randomly picked one chapter for each person while the final book chapter was read by all.
We chose the final chapter as it represents all aspects of our method: it includes music retrieved via movie cues as well as pure emotion based retrieval, apart from being a key chapter in terms of the plot.
This allows us to fairly evaluate the effectiveness of our soundtrack at a book level as reading these chapters takes around an hour for each participant.

All participants are asked to read these chapters consecutively and answer a series of questions about several aspects of the reading experience.
No user data is collected through this process and user consent is collected prior to recording the interviews.
The study was also authorised by the author's institutional ethics committee.

\newpara{Reading application.}
\label{reading_app}
A bottleneck with playlist based approaches for soundtracking books is that they require user-input to transition at appropriate instants.
We resolve this by creating an application that plays music in the background of the displayed text.
Our application loops the music segment infinitely and cross-fades to the next as the reader moves on to the next text segment.
This ensures complete immersiveness and lets participants with \emph{varying} reading speeds truly enjoy the soundtrack. For readers to have a first-hand experience, we package this application for a few chapters at the project page.

\newpara{Participant information.}
A total of 10 individuals (18-22 age range, 6 male, 4 female) volunteered for the study.
Each participant had previously read the \textit{Harry Potter} book without music and happened to be familiar with the movie as well.
We also provided each participant with a small monetary reward as compensation for their time.

\subsection{Findings}

\newpara{The soundtrack improved immersiveness.} 
All participants reported that the soundtrack improved the immersiveness of the reading experience.
This is remarkable as all participants noted that they typically do not listen to music while reading.
Some participants were more receptive to the soundtrack than others and spoke glowingly about the movie-like immersion afforded by the music. 
These participants could recognise that the music played was similar to the score at the relevant movie scene. 
Others, who were positive but less movie-inclined, focused more on how the music \textit{"set the environment"} (P4).
Phrases such as \textit{"set the mood"}, \textit{"intensified the emotion"}, \textit{"fit the vibe"}, and \textit{"enhanced the experience"} were common among such participants.

\begin{displayquote}
\textit{"The music provided insight into the tone of the chapter and [...] beyond imagination, provided a soundtrack to what was read"} (P1)
\end{displayquote}
Surprisingly, some participants reported that the first section of the random chapter that they read first threw them off initially, suggesting that there is an adjustment period to this experience.
These participants reported that they were very comfortable with later sections (\textit{"after the first 10-12 lines"} - P2) once they had gotten into a reading flow.

\newpara{The soundtrack helped visualise the book.}
Some participants could recognise that the music played was similar to the score at the relevant movie scene.
These participants were especially appreciative of the soundtrack and stated that it helped \textit{"visualise the book, with movie like visuals"} (P1). Such participants typically appreciated the music's cohesiveness and were likely referring to the pleasing soundscape laid out by our soundtrack. One participant explicitly pointed out that the soundtrack helped visualise character interactions and that without, it would have just been \textit{"two characters talking"}.

Many were also receptive to the recurring motifs and themes that appeared throughout, such as the central \textit{Harry Potter} theme and stated that it helped imagine the fictional world.
Only one participant, P3, expressed complete disagreement with the music played in a text segment, for ironically the same reason, stating that the signature \textit{Harry Potter} motif distracted them. Barring this, the same person spoke warmly about the remaining soundtrack, suggesting that the overtness of the motif, which permeates culture today, may be subject to individualistic preferences.

\newpara{Music helped focus.}

\begin{displayquote}
    \textit{"I get distracted when reading so it helped me focus on certain parts"} (P5)
\end{displayquote}
When describing the immersiveness of the reading experience, three participants specifically pointed out that the music helped them read continuously, without distractions that are typically present when reading in the absence of a soundtrack. It should be noted that few participants who described an aversion for listening to music during typical recreational reading specifically pointed out that they avoid pop music, suggesting that instrumental music, in general, may be better suited for reading purposes.

\newpara{Moderate repetition is not a concern.}
Most participants noticed repetitive music in some text segments that they read but only brought it up when explicitly asked.
One participant suggested playing a different, similarly composed track instead of repeating the music, but in agreement with the rest of the pool, felt that repetition did not affect the experience.
These participants also said that the repetition wasn't glaringly obvious and that it did not distract or take away from the immersion.
It should be noted that repetition refers to the same track being played on a cross-fade loop while a person reads a text segment, due to variable reading speeds.
Despite this, it is noteworthy that our musical segments are homogeneous enough to be played repeatedly and that the text segments are narratively cohesive to warrant such repetition.
When specifically asked about the diversity of the music played, all participants expressed positive opinions and noted that the tracks kept changing as and when required.

\newpara{Narrative transitions mirrored in music.}
Many participants engaged in a conversation with the authors post-interview about how the soundtrack was built.
All such participants were startled by the fact that the entire process was completely automatic. Their surprise was primarily due to the fact that the music was automatically matched with relevant areas in text and that it transitioned at appropriate narrative points.

\begin{displayquote}
    \textit{"It actually made the experience better as the transition put you in the mood for the expected emotion - from melancholy to sad."} (P9)
\end{displayquote}
When asked specifically about these narrative transitions during the interview, all participants agreed that it was emblematic of emotional or narrative shifts. Since the last chapter was common across all participants, there was strong consensus about the narrative shifts here especially, with participants noting that the music increased in tension as the final hero-villain clash developed and that it eased into more mellow, tender music once it was resolved.

\begin{displayquote}
    \textit{"When the tension built in the plot, the music transitioned to match it."} (P1)
\end{displayquote}
Some participants explicitly appreciated the foreshadowing made possible by the music, as suspenseful music at the start of a segment would precede a similar narrative plotline.
Others simply elaborated on their earlier comments about the immersiveness and re-emphasised the high congruency between text segments and the matched music.

\newpara{Low-arousal music preferred.}
We asked participants to describe the emotional suitability of the music, specifically, and in line with prior answers, and received a favourable response. The few critiques that were present revolved around the energy/arousal of the music played at certain instances. Some participants were dissatisfied with segments that were high in arousal, even though the valence matched, as it broke the immersion. They described how this music distracted them from reading and drew too much attention to the track itself. This is likely an important factor to consider for future work on soundtracking books.

\newpara{Movie-based music cues stronger than emotion-based cues.}
At the end of the interview, participants were asked to describe their favourite and least favoured pieces of music from the two chapters, if any.
The best segments were split between music retrieved via movie cues and those retrieved via emotion matching, with the former being more prominent.
On the other hand, the least favoured segments, often mentioned only after the author's insistence, were typically emotion-based matches that were described in terms of their neutrality.
This finding suggests that our approach can effectively retrieve narrative-specific music and that such a system is perhaps well-suited for book soundtracking, as opposed to pure emotion-based methods.

\section{Conclusion}

We presented a novel system to automatically weave a book-length soundtrack, using the music present in the relevant movie adaptation.
Perceptual validation of the constructed soundtrack for the first book and movie pair of the \textit{Harry Potter} series provided very positive feedback that validated several proposed design decisions and techniques for text, movie, and music processing. 
Participants in our perceptual study were particularly receptive to music that was fetched via the movie and uniformly stated that it improved the immersiveness of the reading experience, and even transported them to the fictional world.

\newpara{Future work.}
While our method has been successful in generating a soundtrack for a book with a movie adaptation, it needs to be modified to make it work on books without adaptations.
Future work can potentially use our approach to investigate common trends across books to determine new cross-narrative rules.
Our approach can also be extended for human-in-the-loop collaborative soundtrack construction applications, though such a use case is beyond the scope of this work.

\section{Acknowledgments}

We thank Siddarth Baasri for his valuable input in analysing motifs present in the \textit{Harry Potter} soundtrack and Saravanan Senthil for illustrating Fig.~\ref{fig:teaser}.

\bibliography{longstrings,ref}

\appendix
\clearpage
\begin{center}
{\Large{\textbf{Supplementary Material}}}
\end{center}

We present some additional details on segmentation (Sec.~\ref{sec:app:segmentation}) and alignment (Sec.~\ref{appendix:book_movie_alignment}).
We also provide additional details of the reading app (Sec.~\ref{sec:app:reading}) and round up the appendix with the questionnaire used in the perceptual study (Sec.~\ref{sec:app:questions}).

\section{Segmenting the Book, Movie, and Music}
\label{sec:app:segmentation}

\subsection{Book Segmentation}

\begin{figure}[t!]
\centering
\includegraphics[width=\linewidth]{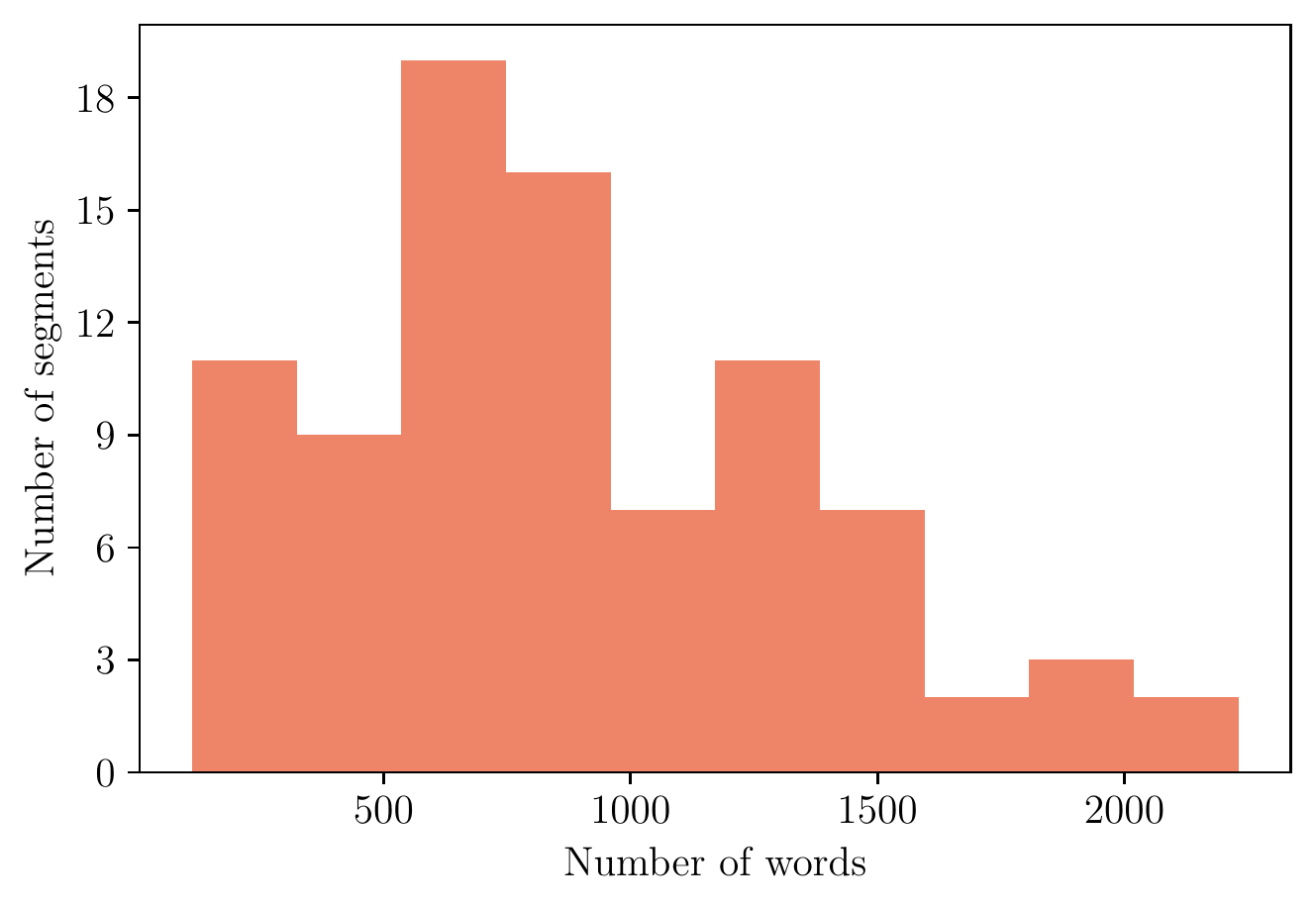}
\caption{Histogram showing the word count distribution for chapter segments.}
\label{fig:hist_segment}
\end{figure}

We segment the book using the hierarchical clustering approach described in Sarfraz~\etal~\cite{sarfraz2021temporally}, using the official implementation%
\footnote{\url{https://github.com/ssarfraz/FINCH-Clustering/tree/master/TW-FINCH}}.
We extract our text features using the \texttt{sentence-transformers} package, which provides several pretrained models for sentence embeddings.
Our chosen model, MPNet~\cite{song2020mpnet}, ranked first on the leaderboard as of writing, on sentence embedding and semantic search tasks.
Specifically, we use the model named \texttt{all-mpnet-base-v2}. For our final segments, we use the segments from the third partition, as it avoided the over-segmentation present at lower partitions while ensuring that segments were homogeneous largely. A histogram of word counts per chapter segment can be seen in Fig.~\ref{fig:hist_segment}.

At one point, we also considered annotating the chapters with text segments to evaluate the efficacy of our approach but quickly came to realise that it would be misleading.
As described in Zehe~\etal~\cite{Zehe2021DetectingSI}, segment boundaries in fiction can be attributed to many, often overlapping reasons.
For instance, a shift in time during a flashback may overlap with a change in location.
Further, depending on the differentiating factor used, segments could overlap; a section divided by an emotion may belong to one when considering time alone.
As such, some of our preliminary evaluations of this approach, using a few annotated chapters, undersold the efficacy of our approach and reported poor F1/accuracy scores. In reality, we find that our text segments are highly plausible and reflective of topical changes in the text, which was also validated by the participants of the perceptual study. We leave a thorough investigation of fiction text segmentation to future work, as it falls outside the scope of this paper.

\begin{figure}[t]
\centering
\includegraphics[width=\linewidth]{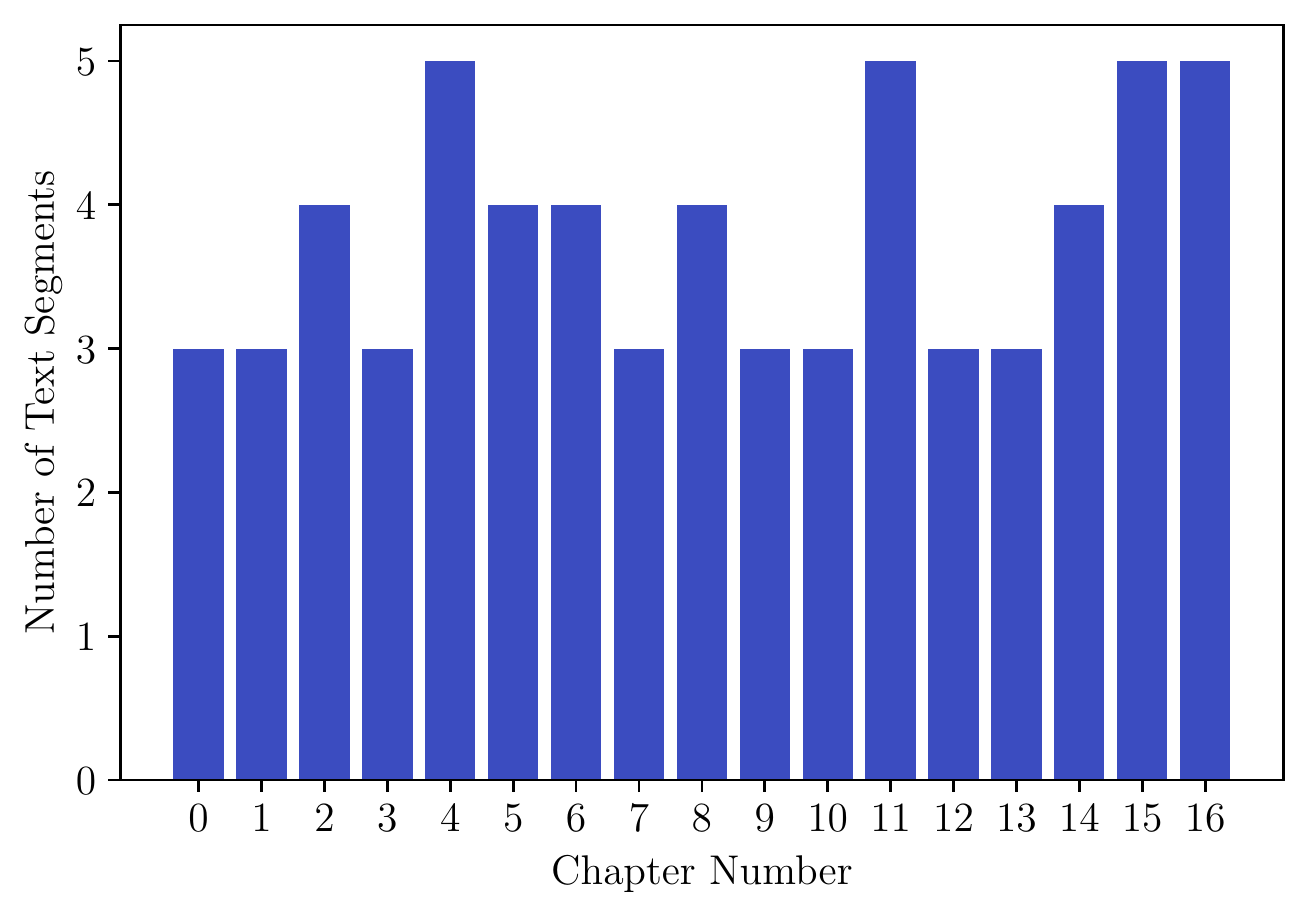}
\caption{Number of chunks per chapter of \textit{Harry Potter} on using TextTiling~\cite{Hearst1997TextTS}.}  
\label{fig:text_tiling}
\end{figure}

\begin{figure}[t]
\centering
\includegraphics[width=\linewidth]{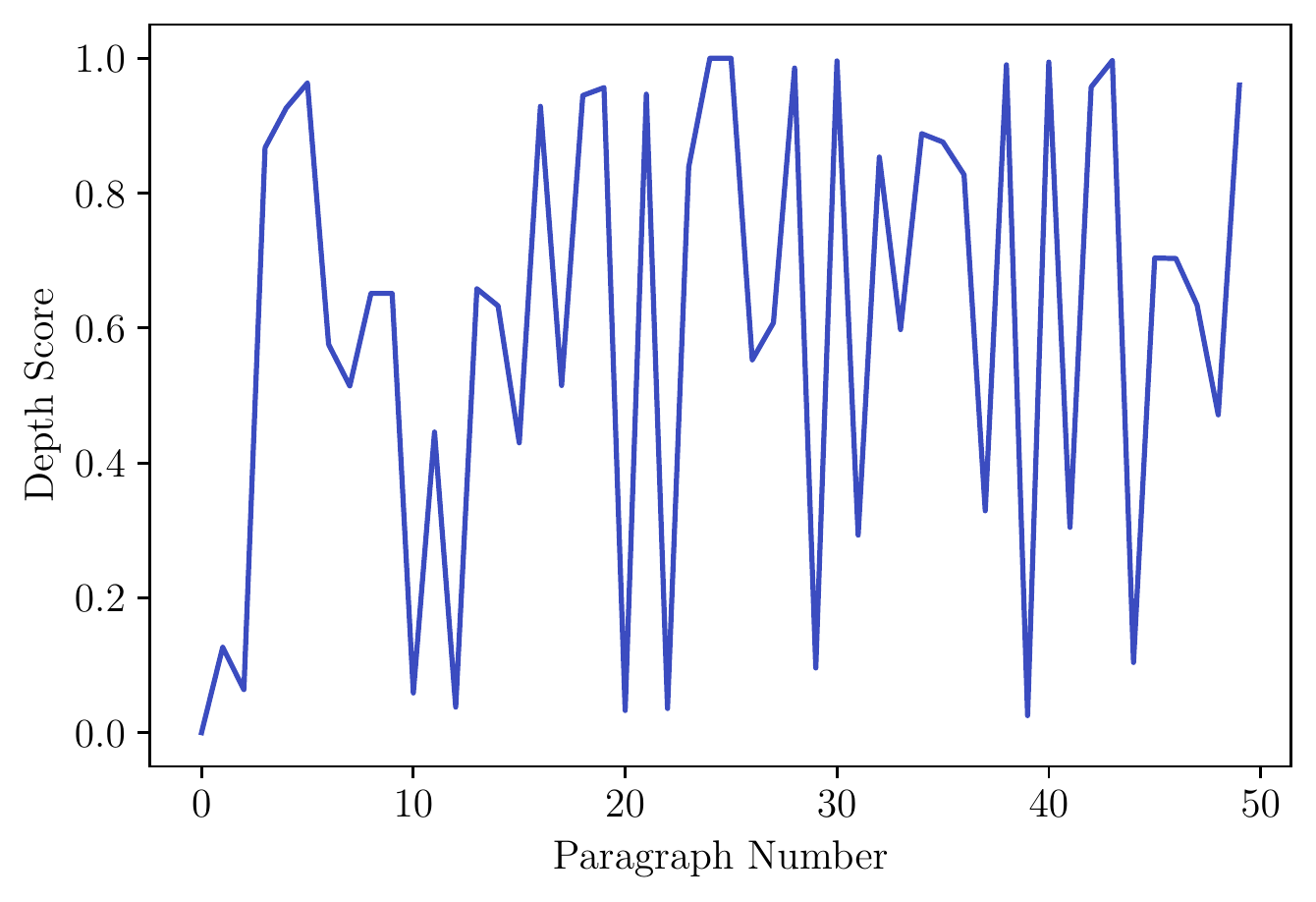}
\caption{Depth score of various segment boundaries, as computed by TopicTiling~\cite{Riedl_topictil}, for a single chapter of \textit{Harry Potter}. Higher values suggest the presence of a boundary.}
\label{fig:topictil}
\end{figure}

We also tried TextTiling~\cite{Hearst1997TextTS}, using the NLTK implementation%
\footnote{\url{https://www.nltk.org/_modules/nltk/tokenize/texttiling.html}}, 
but found that it uniformly partitioned the text typically, with every chapter containing 3-5 segments regardless of content (see Fig.~\ref{fig:text_tiling}). In addition, we tried TopicTiling~\cite{Riedl_topictil} which uses Latent Dirichlet Allocation topic models to segment text, using its official implementation%
\footnote{\url{https://github.com/riedlma/topictiling}}.
This produces a series of depth score, indicating the likelihood of having a segment boundary at that instant. As seen in Fig.~\ref{fig:topictil}, TopicTiling seemed to over-segment the text, and produced several short segments.
Both these methods have also been shown to be perform poorly on a German variant of the fiction segmentation task~\cite{Zehe2021DetectingSI}.

\begin{figure*}[b!]
\centering
\includegraphics[width=\linewidth]{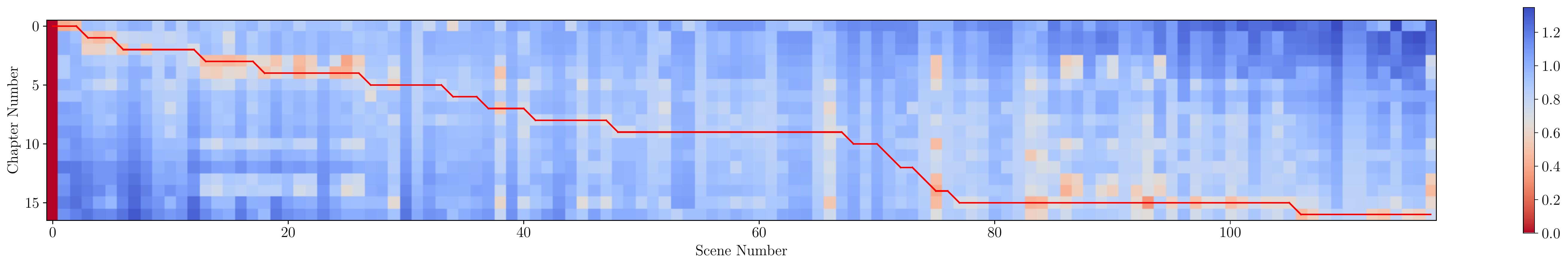}
\caption{Visualisation of the chapter-scene (17 chapters, 120 scenes) similarity scores computed for \textit{Harry Potter}. The red line on the matrix shows the shortest path on the graph, which aligns the book and movie.}  
\label{fig:b2m_matrix}
\end{figure*}

\subsection{Movie Segmentation}

We segment the movie using the shot threading approach described in Tapaswi~\etal~\cite{tapaswi_storygraph}, with its official \texttt{MATLAB} implementation%
\footnote{\url{https://github.com/makarandtapaswi/Video_ShotThread_SceneDetect/}}.
As this approach requires pre-computed video shots, we use an open-source package\footnote{\url{https://github.com/makarandtapaswi/shotdetection/}} to obtain shots.
For \textit{Harry Potter}, we obtain 2,525 shots.
We also tried \textit{PySceneDetect} for shot/scene detection but found that it resulted in over-segmentation, which is undesirable when trying to obtain high-confidence matches between the text and movie.

\subsection{Music Segmentation}

Our choice of novelty based segmentation is a reflection of how tracks undergo emotional and tonal shifts. After multiple rounds of testing, we settled on a kernel width of 64 for this, as it produced perceptually cohesive segments that were sufficiently long.

\section{Aligning the Book, Movie, and Music}
\label{appendix:book_movie_alignment}

\subsection{Chapter-Scene Coarse Alignment}

In order to align book chapters with movie scenes, we largely reproduce the approach described in Tapaswi~\etal~\cite{TapaswiB2M}.
Their approach parses each chapter and video scene to collect a collection of dialogues and characters, identified by POS tagging on the text/script/subtitles.
A similarity score is then computed between the two on two counts - the similarity of character histograms and length of the longest common subsequence (LCS) between movie and book dialogues.
We refer the reader to the original paper for more details on its implementation.

Our implementation did however, differ in a few aspects. We align the subtitles with the movie transcript (obtained online\footnote{\url{https://warnerbros.fandom.com/wiki/Harry_Potter_and_the_Philosopher\%27s_Stone/Transcript}}) using Dynamic Time Warping with LCS as the distance function, to obtain speaker identities for all dialogues, as the subtitles did not provide this. The original paper used facetracks instead.

To obtain character names, we process the entire book using BookNLP\footnote{\url{https://github.com/booknlp/booknlp}}, which detects quotes and performs speaker attribution. We only count those characters who have dialogue, unlike in POS Tagging, as they are more likely to be seen in the movie. We use the character names obtained here for our histogram. We automatically match characters names in the movie and book based on LCS, to obtain a common space for our character histogram.

Next, we compute an inverse character frequency to prioritise rare characters who can provide strong alignment cues and scale our character histogram accordingly.

We set $\alpha=1$ (Eq.~6 in Tapaswi~\etal~\cite{TapaswiB2M}) and use equal weights for the character and dialogue similarity scores.
The resulting alignment is visualised in Fig.~\ref{fig:b2m_matrix}.

\subsection{Movie Audio-Soundtrack Alignment}

We use Shazam API\footnote{\url{https://github.com/Numenorean/ShazamAPI}} in \textit{Python} to identify music from the movie audio.
We also tried an open-source implementation of audio fingerprinting and music identification, DejaVu\footnote{\url{https://github.com/worldveil/dejavu/}}, by registering the soundtrack album in a custom database. However, this approach failed to effectively recognise music in the presence of noise, though different parameters may yield better results.

\section{Reading Application}
\label{sec:app:reading}

In order to facilitate a smooth reading experience, we built a custom reading app to ensure that readers with \emph{varying} reading speeds could enjoy our soundtrack.  We built our reading app using React\footnote{\url{https://reactjs.org/}} and used simple state manipulation to change the track played based on scroll position. We dynamically cross-fade songs based on scroll position using the Web Audio API.

\section{Study Questionnaire}
\label{sec:app:questions}

As part of the perceptual study, we ask our volunteers several questions in a semi-structured oral interview:
\begin{enumerate}[itemsep=-1pt]
\item What is your general feedback with what you've read/heard?
\item Compare your reading experience with music to one without.
\item How was the music placed with respect to the text / How aligned is the text and music?
\item Would you say that the the music is repetitive? Describe your thoughts with respect to the diversity of music played as well as how often a single song plays.
\item Different music is played at different points in the book, did this have any effect on your reading experience? 
\item How did the music transitions affect your reading experience?
\item How did the music transitions align with changes in text? 
\item Did the soundtrack reflect the emotion present in the text? How well did it do? 
\item Were there any instances that you particularly liked? Were there any instances that you didn't like?
\end{enumerate}

\end{document}